\documentstyle[prl,aps,multicol]{revtex}

\begin{document}
\draft

\title{ Electronic origin of magnetic and orbital ordering in insulating
                             LaMnO$_3$ }

\author { Louis Felix Feiner\cite{LFF} }
\address{ Philips Research Laboratories, Prof. Holstlaan 4,
          NL-5656 AA Eindhoven, The Netherlands, and \\
          Institute for Theoretical Physics, Utrecht University,
          Princetonplein 5, NL-3584 CC Utrecht, The Netherlands }
\author { Andrzej M. Ole\'{s}\cite{AMO} }
\address{ Max-Planck-Institut f\"ur Festk\"orperforschung,
          Heisenbergstrasse 1, D-70569 Stuttgart,
          Federal Republic of Germany }

\date{November 19, 1997; revised May 1, 1998}
\maketitle

\begin{abstract}
We derive a spin-orbital model for insulating LaMnO$_3$ which fulfills
the $SU(2)$ symmetry of $S=2$ spins at Mn$^{3+}$ ions.
It includes the complete $e_g$ and $t_{2g}$ superexchange which follows
from a realistic Mn$^{2+}$ multiplet structure in cubic site symmetry,
and the Jahn-Teller induced orbital interactions. We show that the magnetic
ordering observed in LaMnO$_3$ is stabilized by a purely electronic
mechanism due to the $e_g$-superexchange alone, and provide for the first
time a quantitative explanation of the observed transition temperature and
the anisotropic exchange interactions.
\end{abstract}
\pacs{PACS numbers: 71.27.+a, 75.10.-b, 75.30.Et, 75.70.Pa.}

\begin{multicols}{2} 

The fascinating properties of doped manganites R$_{1-x}$A$_x$MnO$_3$,
where R is a rare earth element, and A is a divalent element, were
discovered almost half a century ago \cite{Jon50}, but the various phase
transitions occurring under doping and in particular the phenomenon of
`colossal magnetoresistance' (CMR) are still not fully understood.
The phase diagrams of La$_{1-x}$(Ca,Sr)$_x$MnO$_3$ \cite{Ram97}
show a complex interplay between magnetic, charge, and structural order,
so that all these ordering phenomena may affect CMR at
least indirectly. It is therefore important to obtain first of all a
full understanding of the mechanism(s) stabilizing the observed order in
the {\em undoped\/} insulating parent compound LaMnO$_3$.
This will be an essential element in putting together a satisfactory
description of the more complicated {\em doped\/} compounds, and
recognizing which mechanism(s) other than or in
addition to double exchange \cite{And55} might be
responsible for CMR \cite{Mil95}.

In this Letter we therefore reconsider the problem of the microscopic
origin of the experimentally observed type of antiferromagnetic (AF)
order in LaMnO$_3$ \cite{Wol55}, which consists of ferromagnetic (FM)
planes ordered antiferromagnetically in the third direction (A-AF phase).
As the magnetic order in LaMnO$_3$ couples to orbital order
\cite{Goo55}, one possible explanation might be the occurrence of a
cooperative Jahn-Teller (JT) effect \cite{Mil96a} which induces a
particular order of the singly occupied $e_g$ orbitals \cite{Kug73}.
However, while the JT effect plays a crucial role in charge transport
\cite{Mil96b}, we show here that a purely {\em electronic mechanism\/}
drives orbital and magnetic ordering in the manganites near the
Mott-Hubbard transition \cite{Var96}.

The local Coulomb interaction $U$ is the dominating energy scale in late
transition metal oxides. If partly filled orbitals are degenerate, as in
KCuF$_3$ or in LaMnO$_3$, this leads to an effective low-energy Hamiltonian,
where {\it spin and orbital\/} degrees of freedom are interrelated
\cite{Kug73,Cas78}. In the simplest case of $d^9$ ions in KCuF$_3$, such a
model describes spins $S=1/2$ of $e_g$ holes coupled to the discrete orbital
variables. Finite Hund's rule exchange $J_H$ removes the classical degeneracy
of magnetically ordered phases \cite{Kug73,Fei97}, and stabilizes the
A-AF phase
in conjunction with the particular orbital order observed in KCuF$_3$.
Here we show that a similar state follows from a {\em realistic
$S=2$ spin-orbital model\/} for the $d^4$ ions in
LaMnO$_3$. We include also the $t_{2g}$ superexchange and the JT
interaction and show that these, while unessential qualitatively,
are important for a quantitative understanding.

The {\it superexchange between total spins\/} $S=2$ at the $d^4$ Mn$^{3+}$
ions originates in the large-$U$ regime from virtual ($e_g$ or $t_{2g}$)
excitations, $d_i^4d_j^4\rightleftharpoons d_i^3d_j^5$. A simplified
approach proposed recently by Ishihara {\it et al.} \cite{Ish96}
emphasizes the role of orbitals but violates the $SU(2)$ spin
symmetry, and involves a Kondo coupling between $e_g$ and $t_{2g}$ spins,
which by itself is not a faithful approximation to the multiplet
structure.
The latter objection applies also to the model proposed by Shiina
{\it et al.} \cite{Shi97}. In contrast,
the spin-orbital model presented below
follows from the full multiplet structure of the Mn ions in
octahedral symmetry, both in the $d^4$ ($t_{2g}^3e_g$) configuration of the
Mn$^{3+}$ ground state and in the $d^3$ and $d^5$ virtually excited states.

So we consider a spin-orbital model for the manganites,
\begin{equation}
\label{full}
H=H_{e}+H_{t}+H_{\rm JT}+H_{\tau},
\end{equation}
which includes superexchange terms due to $e_g$ ($H_e$) and $t_{2g}$ ($H_t$)
excitations, JT interaction ($H_{\rm JT}$), and a low-symmetry crystal field
($H_{\tau}$). Our starting point is that each Mn$^{3+}$ ($d^4$) ion is in the
strong-field ($t_{2g}^3e_g$) Hund's rule ground state, i.e., the high-spin
($S=2$) orbital doublet $^5\!E$. First, we analyze the strongest channel of
superexchange, which originates in the hopping of an $e_g$ electron from site
$i$ to its neighbor $j$. When we consider a bond oriented along the cubic
$c$-axis, only a $3z^2-r^2$ electron can hop and four $d^5$ states may
be reached: the high-spin $^6\!\!A_1$ state ($S=5/2$), and the lower-spin
($S=3/2$) $^4\!\!A_1$, $^4\!E$, and $^4\!\!A_2$ states (Fig. \ref{virtual}).
The $d_i^4d_j^4\rightleftharpoons d_i^3(t_{2g}^3)d_j^5(t_{2g}^3e_g^2)$
excitation
energies require for their description in principle {\em all three\/} Racah
parameters, $A$, $B$ and $C$ \cite{Gri71}:
$\varepsilon(^6\!\!A_1)=A-8B$,  $\varepsilon(^4\!\!A_1)=A+ 2B+5C$,
$\varepsilon(^4\!E  )\simeq A+6B+5C$ \cite{notee},
$\varepsilon(^4\!\!A_2)=A+14B+7C$.
In view of the realistic values of $B=0.107$ and $C=0.477$ eV for Mn$^{2+}$
($d^5$) ions \cite{Boc92}, one may use an approximate relation $C\simeq 4B$,
and write the excitation energies in terms of Coulomb, $U\equiv A+2B+5C$,
and Hund's exchange, $J_H\equiv 2B+C$, parameters:
$\varepsilon(^6\!\!A_1) = U - 5 J_H$,
$\varepsilon(^4\!\!A_1) = U$,
$\varepsilon(^4\!E\,) = U + \frac {2}{3} J_H$,
$\varepsilon(^4\!\!A_2) = U + \frac{10}{3} J_H$. Using
the spin algebra (Clebsch-Gordon coefficients) and the reduction of product
representations in cubic site symmetry \cite{Gri71} for the intermediate
states, and making a rotation of the terms derived for a bond
$\langle ij\rangle\parallel c$ with respect to the cubic axes,
one finds a compact expression,
\begin{eqnarray}
\label{egterm}
H_{e}&=&\frac{1}{16}\sum_{\langle ij\rangle}\left\{
 - \frac{8}{5} \frac{t^2}{\varepsilon(^6\!\!A_1)}
   \left(\vec{S}_i\cdot\vec{S}_j+6\right)
   {\cal P}_{\langle ij\rangle}^{\zeta\xi}\right.              \nonumber \\
&+& \left. \left[\frac{t^2}{\varepsilon(^4\!E)}
   + \frac{3}{5}\frac{t^2}{\varepsilon(^4\!\!A_1)} \right]
   \left(\vec{S}_i\cdot\vec{S}_j-4\right)
   {\cal P}_{\langle ij\rangle}^{\zeta\xi}\right.              \nonumber \\
&+&\left. \left[ \frac{t^2}{\varepsilon(^4\!E)}
   + \frac{t^2}{\varepsilon(^4\!\!A_2)} \right]
   \left(\vec{S}_i\cdot\vec{S}_j-4\right)
   {\cal P}_{\langle ij\rangle}^{\zeta\zeta}\right\},
\end{eqnarray}
where $t$ is the hopping element along the $c$-axis,
and ${\cal P}_{\langle ij\rangle}^{\alpha\beta}$ are projection operators
for each bond $\langle ij\rangle$,
\begin{equation}
\label{porbit}
{\cal P}_{\langle ij\rangle}^{\zeta\xi}=
 P_{i\zeta}P_{j  \xi}+P_{i  \xi}P_{j\zeta}, \hskip .7cm
{\cal P}_{\langle ij\rangle}^{\zeta\zeta}=2P_{i\zeta}P_{j\zeta},
\end{equation}
projecting on the orbital states, being either parallel
to the bond direction on one site
($P_{i\zeta}=\frac{1}{2}-\tau^{\alpha}_i$)
and perpendicular on the other
($P_{j  \xi}=\frac{1}{2}+\tau^{\alpha}_j$),
or parallel on both sites. They are represented by the orbital operators
$\tau^{\alpha}_i$ associated with the three cubic axes ($\alpha=a$, $b$,
or $c$),
\begin{equation}
\label{orbop}
\tau^{a(b)}_i = \frac{1}{4}( -\sigma^z_i\pm\sqrt{3}\sigma^x_i ), \hskip .7cm
\tau^c_i = \frac{1}{2} \sigma^z_i,
\end{equation}
where the $\sigma$'s are Pauli matrices acting on the orbital
pseudo-spins
$|x\rangle ={\scriptsize\left( \begin{array}{c} 1\\ 0\end{array}\right)},\;
 |z\rangle ={\scriptsize\left( \begin{array}{c} 0\\ 1\end{array}\right)}$,
and the orbitals transform as $|x\rangle \propto x^2-y^2$ and
$|z\rangle \propto (3z^2-r^2)/\sqrt{3}$.

The spin operators $\vec{S}_i$ in Eq. (\ref{egterm}) are $S=2$ spins, but
otherwise $H_e$ resembles the spin-orbital model for $d^9$ ions in the
cuprates \cite{Fei97}. Both models contain superexchangelike couplings
between spin and orbital degrees of freedom. The orbital sector carries
a discrete cubic symmetry, and is {\it identical\/} in both cases,
while the {\it spin problem fulfills the $SU(2)$ symmetry\/}, and different
representations apply for the manganites ($S=2$) and for the cuprates
($S=1/2$). We emphasize that the Hamiltonian $H_e$ {\em is not equivalent\/}
to that of Ref. \cite{Ish96} in any nontrivial limit. A common
feature is that FM interactions are enhanced due to the lowest excited
$^6\!\!A_1$ state, but the dependence of the magnetic interactions on $J_H$
is quite different, and it gives a different answer concerning the stability
of the A-AF phase.
The balance between AF and FM interactions is also different from that
in Ref. \cite{Shi97} due to the multiplet structure of Mn$^{2+}$.

A similar derivation gives the $t_{2g}$ superexchange \cite{noteanis},
\begin{equation}
\label{t2gterm}
H_{t} = \frac{1}{4}J_t\sum_{\langle ij\rangle}
           \left(\vec{S}_i\cdot\vec{S}_j-4\right) ,
\end{equation}
where $J_t=(J_{11}+J_{22}+J_{12}+J_{21})/4$. The exchange elements,
$J_{mn}=t_{\pi}^2/\varepsilon(^4T_m,^4\!T_n)$, where $t_{\pi}=t/3$ is the
hopping between the $t_{2g}$ orbitals, result from local
$d_i^4d_j^4 \rightleftharpoons d_i^5(t_{2g}^4e_g)d_j^3(t_{2g}^2e_g)$
excitations within a $\langle ij\rangle$ bond, with energies
$\varepsilon(^4T_1,^4\!T_1)\simeq U+8J_H/3$,
$\varepsilon(^4T_1,^4\!T_2)\simeq U+2J_H/3$,
$\varepsilon(^4T_2,^4\!T_1)\simeq U+4J_H$,
$\varepsilon(^4T_2,^4\!T_2)\simeq U+2J_H$,
where $^4T_m$ ($^4T_n$) stands for the symmetry of $d_i^5$ ($d_j^3$)
excited configurations, respectively.

The manganite model (\ref{full}) is completed by the JT term which leads
to static distortions and mixes $e_g$ orbitals \cite{Mil96a},
\begin{equation}
\label{jtterm}
H_{\rm JT}=\kappa\sum_{\langle ij\rangle}\left(
           {\cal P}_{\langle ij\rangle}^{\zeta\zeta}
         -2{\cal P}_{\langle ij\rangle}^{\zeta\xi  }
          +{\cal P}_{\langle ij\rangle}^{\xi  \xi  }\right),
\end{equation}
with ${\cal P}_{\langle ij\rangle}^{\xi\xi}=2P_{i\xi}P_{j\xi}$, and by the
tetragonal crystal field,
\begin{equation}
\label{ezh}
H_{\tau}=-E_z\sum_{i}\tau_{i}^c.
\end{equation}

The strength of $e_g$ and $t_g$ superexchange can be estimated fairly
accurately from the basic electronic parameters for the Mn ion as determined
from spectroscopy \cite{Boc92,Miz95} with an estimated accuracy of $\sim
10\%$. We thus use $U=7.3$ eV and $J_H=0.69$ eV, and taking into account
that the Mn-Mn hopping occurs via the bridging oxygen, $t=0.41$ eV as
follows from $t=t_{pd}^2/\Delta$ with Mn-O hopping $t_{pd}=1.5$ eV and
charge transfer energy $\Delta=5.5$ eV \cite{notepar}.
This yields $J=t^2/U=23$ meV and $J_t=2.1$ meV.
The accuracy of these parameters may be appreciated from the resulting
prediction for the N\'eel temperature of CaMnO$_3$, where a similar
derivation gives
$\hat{H}_t\sim 2\hat{J}_t(\frac{4}{9}\vec{S}_i\cdot\vec{S}_j-1)$ in terms
of Mn$^{4+}$ spins $S=3/2$ and $\hat{J}_t\approx J_t(1+J_H/U)$. With our 
present estimates we obtain $\hat{J}_t=2.3$ meV and thus $T_N=124$ K, in
excellent agreement with the experimental value $T_N=110$ K \cite{Wol55}.

When considering the manganite ($d^4$) model (\ref{full}), it is
instructive to
treat $J_H$ and $E_z$ as freely variable parameters in order to
appreciate the physical consequences of Hund's rule multiplet
splitting and orbital degeneracy. The cuprate ($d^9$) model exhibits
symmetry breaking into classical states with simultaneous spin and
orbital order \cite{Kug73,Fei97}, and similar behavior is expected
here \cite{notesl}.
We have considered classical phases with two and four sublattices, and
mixed orbitals (MO), $|i\mu\sigma\rangle=
\cos\theta_i|ix\sigma\rangle+\sin\theta_i|iz\sigma\rangle$.
The mean-field (MF) phase diagram of the $e_g$-part of model (\ref{full}),
$H=H_e+H_{\tau}$, at $T=0$ is similar to that of the cuprate spin-orbital
model \cite{Fei97}: at large positive (negative) $E_z$, one finds AF phases
with either $|x\rangle$ (AFxx) or $|z\rangle$ (AFzz) orbitals occupied,
while MO phases with orbitals alternating between the sublattices
($\theta_i=\pm\theta$, with $\cos 2\theta <0$) are favored by increasing
$J_H$. If $E_z<0$ the spin order is FM (AF) in the $(a,b)$ planes (along the
$c$-axis) in the MO{\footnotesize FFA} phase, while at $E_z>0$ two similar
phases, MO{\footnotesize AFF} and MO{\footnotesize FAF}, are degenerate.
For the parameters appropriate for LaMnO$_3$ ($J_H/U\simeq 0.095$) one
finds a MO{\footnotesize FFA}/MO{\footnotesize AFF} ground state, i.e.,
{\em A-AF magnetic order\/}, while a FM (MO{\footnotesize FFF}) phase is
found only at $J_H/U>0.12$. The region of stability of the A-AF phase is
modified by $t_{2g}$-superexchange [Fig. \ref{mfa}(a)], but this change is
small as $J_t\ll J$. Thus, the observed A-AF magnetic order in LaMnO$_3$
is caused by the {\em orbital dependence\/} of the $e_g$-superexchange
and not by competition between FM $e_g$- and AF $t_{2g}$-superexchange
as proposed in Ref.~\onlinecite{Ish96} (where an unrealistically large
$J_t$ was used), supporting the qualitative results of Ref. \cite{Shi97}.

Although the MF phase diagrams are modified significantly by JT coupling
[Fig. \ref{mfa}(b)], the A-AF phase survives around $J_H/U=0.095$. In fact,
the JT interaction (\ref{jtterm}) by itself enforces alternating orbitals
with $\cos 2\theta=0$, which favors AF spin order, thus stabilizing at small
$J_H/U$ the MO{\footnotesize AAA} phase, promoted further by finite $J_t$.
But at larger $J_H/U$, even though the JT interaction sets the stage
by inducing orbital order as such, the actual magnetic (A-AF) and orbital
($\cos 2\theta\neq 0$) order are {\em entirely due\/} to the
$e_g$-superexchange interactions (\ref{egterm}).

Finite temperature behavior was investigated in MF approximation,
with $\langle \vec{\sigma} \rangle$,
$\langle \vec{S} \rangle$, and
$\langle \vec{\sigma}\vec{S} \rangle$ constituting independent order
parameters \cite{Dit80}. As the largest interaction is in the pure
orbital ($\vec{\sigma}$) channel, one may estimate the JT coupling
$\kappa$ from the temperature of the structural transition,
$T^{exp}_s\approx 750$ K. The electronic interactions contribute $\simeq
440$ K (Fig. \ref{tc}), and the rest, $6\kappa\simeq 760$ K, comes from
the JT term \cite{notetc}. Thus $\kappa\simeq 11$ meV, and we have
adopted the representative value $\kappa/J=0.5$. We then calculated the
temperatures $T_c$ for the possible magnetic transitions (Fig.
\ref{tc}), taking into account that orbital order with $\cos 2\theta=0$
already exists below $T_s$, and calculating selfconsistently the
corresponding order parameter $\langle\vec{\sigma}\rangle$ at finite
$T$. The spin order sets in simultaneously with a modification of
orbital ordering towards $\cos 2\theta\neq 0$. We find that the
preexisting structural transition plays an important role at finite $T$
and reduces the magnetic transition temperature, being otherwise
$T_c\simeq J$ \cite{Shi97}. The results are consistent with the phase
diagrams at $T=0$ (Fig. \ref{mfa}), as the magnetic transition
corresponds to the same order as found at $T=0$. For the A-AF
(MO{\footnotesize FFA}) phase we find $T_{c}\simeq 106$ K \cite{notetc},
in reasonable agreement with the experimental value of 136 K \cite{Kaw96}.

The magnetic interactions in the A-AF (MO{\footnotesize FFA}) phase may
be found using averages
$\langle{\cal P}_{\langle ij\rangle}^{\alpha\beta}\rangle$ of the orbital
projection operators (\ref{porbit}) at $E_z=0$. They are FM in the $(a,b)$
planes ($J_{(a,b)}$), and AF in the $c$-direction ($J_c$) (Fig. \ref{jab&jc}).
Both large $J_H/U$ and $\kappa=0.5J$ play a decisive role in determining
the actual composition of the orbitals, and we find $J_{(a,b)}=-1.15$ and
$J_c=0.88$ meV, somewhat higher than the experimental -0.83 and
0.58 meV \cite{Hir96}. However, their ratio, $J_c/|J_{(a,b)}|=0.77$, agrees
very well with the experimental value of $0.7-0.72$ \cite{Hir96};
in contrast, it would amount to 1.04 for $\kappa=0$, and to 2.25
if in addition the orbitals were chosen to satisfy $\cos 2\theta=-0.5$.

Summarizing, a coherent overall picture has been obtained for LaMnO$_3$,
which is even quantitatively satisfactory.
It includes simultaneously the full ($e_g$ and
$t_{2g}$) superexchange and the JT effect, and shows that the orbital
dependence of the $e_g$-superexchange, a {\em purely electronic mechanism\/},
is responsible for the observed A-AF order. We emphasize that no fitting of
parameters was needed, and the used values of $J_H$, $U$, and $t$, known with
an accuracy of $\sim 10\%$, allowed to deduce the value of the JT coupling
$\kappa$, and gave $T_c$, $J_{(a,b)}$ and $J_c$ within $30\%$ from the
experimental values. We thus believe that Hamiltonian (\ref{full})
provides a realistic starting point for understanding
how the delicate balance of magnetic and orbital interactions
in LaMnO$_3$ is affected by doping, leading to a change of magnetic
order and to the possible onset of an orbital liquid state \cite{Nag97}.

We thank P. Horsch, J. Zaanen, D.I. Khomskii, H.J.F. Knops and H. Shiba
for valuable discussions, and acknowledge the support by the Committee
of Scientific Research (KBN) of Poland, Project No.~2 P03B 175 14.

\narrowtext

\begin{figure}
\caption
{Virtual $d_i^4d_j^4\rightarrow d_i^3d_j^5$ excitations
 which generate effective interactions for a bond $(ij)\parallel c$-axis:
 (a) for one $|x\rangle$ and one $|z\rangle$ electron, and
 (b) for two $|z\rangle$ electrons. }
\label{virtual}
\end{figure}

\begin{figure}
\caption
{Classical phase diagram of the manganite model (\protect{\ref{full}}):
 (a) no JT effect ($\kappa=0$), $J_t=0$      (full lines) and
                                $J_t=0.092J$ (dashed lines),
     with the AFxx and AFzz phases separated by a MOAAA phase;
 (b) including JT effect ($\kappa=0.5J$), $J_t=0$      (dashed lines) and
                                          $J_t=0.092J$ (full lines). }
\label{mfa}
\end{figure}

\begin{figure}
\caption
{Magnetic transition temperatures $T_c$ ($J=23$ meV, $E_z=0$, $J_t=0.092J$,
 $\kappa=0.5J$) for:
 MOAAA (dashed line), MOFFF (long-dashed line), and MOFFA (full line)
 phases, and $T_s$ for the structural (MO) phase transition at $\kappa=0$
 (squares). The dotted line indicates realistic $J_H/U=0.095$. }
\label{tc}
\end{figure}

\begin{figure}
\caption
{Exchange interactions $J_{ab}$ and $J_c$ in the
 ground state for increasing $J_H/U$, for $J=23$ meV and:
 $J_t=0$,      $\kappa=0$    (dashed lines),
 $J_t=0.092J$, $\kappa=0$    (long-dashed lines), and
 $J_t=0.092J$, $\kappa=0.5J$ (full lines).
 The inset shows $\cos 2\theta$. }
\label{jab&jc}
\end{figure}

\end{multicols} 


\begin{references}

\bibitem[*]{LFF} E-mail: feiner@natlab.research.philips.com   

\bibitem[\dagger]{AMO} Permanent address: Institute of Physics, Jagellonian
                 University, Reymonta 4, PL-30059 Krak\'ow, Poland.

\bibitem{Jon50} G. Jonker and J. van Santen,
                   Physica {\bf 16}, 337 (1950).

\bibitem{Ram97} A. P. Ramirez,
                   J. Phys. Cond. Matter. {\bf 9}, 8171 (1997).

\bibitem{And55} P. W. Anderson and H. Hasegawa,
                   Phys. Rev. {\bf 100}, 675 (1955);
                P.-G. de Gennes,
                   Phys. Rev. {\bf 118}, 141 (1960).

\bibitem{Mil95} A. J. Millis {\it et al.},
                   Phys. Rev. Lett. {\bf 74}, 5144 (1995).

\bibitem{Wol55} E. O. Wollan and W. C. Koehler,
                   Phys. Rev. {\bf 100}, 545 (1955).

\bibitem{Goo55} J. B. Goodenough,
                   Phys. Rev. {\bf 100}, 564 (1955).

\bibitem{Mil96a} B. Halperin and R. Englman,
                   Phys. Rev. B {\bf 3}, 1698 (1971);
                A. J. Millis,
                   Phys. Rev. B {\bf 53}, 8434 (1996).

\bibitem{Kug73} K. I. Kugel and D. I. Khomskii,
                   Sov. Phys. JETP {\bf 37}, 725 (1973).

\bibitem{Mil96b} A. J. Millis {\it et al.},
                   Phys. Rev. B {\bf 54}, 5389 and 5405 (1996).

\bibitem{Var96} C. M. Varma,
                   Phys. Rev. B {\bf 54}, 7328 (1996).

\bibitem{Cas78} C. Castellani, C. R. Natoli, and J. Ranninger,
                   Phys. Rev. B {\bf 18}, 4945 and 4967 (1978);
                T. M. Rice,
                   in {\it Spectroscopy of Mott Insulators and Correlated
                   Metals}, edited by A. Fujimori and Y. Tokura
                   (Springer Verlag, Berlin, 1995).

\bibitem{Fei97} L. F. Feiner, A. M. Ole\'s, and J. Zaanen,
                   \prl {\bf 78}, 2799 (1997).

\bibitem{Ish96} S. Ishihara, J. Inoue, and S. Maekawa,
                   Physica C {\bf 263}, 130 (1996);
                   Phys. Rev. B {\bf 55}, 8280 (1997).

\bibitem{Shi97} R. Shiina, T. Nishitani, and H. Shiba,
                   J. Phys. Soc. Jpn. {\bf 66}, 3159 (1997).

\bibitem{Gri71} J. S. Griffith,
                   {\it The Theory of Transition Metal Ions}
                   (Cambridge University Press, Cambridge, 1971).

\bibitem{notee} Only a particular linear combination of two $^4\!E$ states
                   can be reached by hopping which has average energy
                   $\varepsilon(^4\!E)$.


\bibitem{Boc92} A. E. Bocquet {\it et al.},
                   Phys. Rev. B {\bf 46}, 3771 (1992);
                J. Zaanen and G. A. Sawatzky,
                   J. Sol. State Chem. {\bf 88}, 8 (1990).

\bibitem{noteanis} We neglected here smaller anisotropic orbital-dependent
                   terms $\sim J_H/U$ that cannot affect the results.

\bibitem{Miz95} T. Mizokawa and A. Fujimori,
                   Phys. Rev. B {\bf 51}, 12\ 880 (1995);
                   Phys. Rev. B {\bf 54}, 5368 (1996).

\bibitem{notepar}  Note that $U$ and $\Delta$ are defined with respect
                   to the Mn$^{2+}$ $^4\!A_1$ state, in contrast to
                   other conventions \protect{\cite{Boc92,Miz95}}.

\bibitem{notesl} Different symmetry-broken states are degenerate
                 at $J_H=E_z=0$ , as in the $d^9$ model
                 \protect{\cite{Fei97}}, revealing a high frustration of
                 magnetic interactions which might lead to a spin liquid
                 with strong orbital correlations. Such quantum behavior
                 is suppressed for the parameters of LaMnO$_3$ with
                 sizable $J_H/U$, and because of the large $S=2$ spins.

\bibitem{Dit80} R. G. Ditzian {\it et al.},
                   Phys. Rev. B {\bf 22}, 2542 (1980).

\bibitem{notetc} The MF values typically overestimate transition temperatures
                by a factor close to 1.6; it reduces the used MF value of
                $T_{s}=1200$ K to the observed $T_{s}^{exp}\simeq 750$ K.

\bibitem{Kaw96} H. Kawano {\it et al.},
                   Phys. Rev. B {\bf 53}, R14\ 709 (1996).

\bibitem{Hir96} K. Hirota {\it et al.},
                   J. Phys. Soc. Jap. {\bf 65}, 3736 (1996);
                F. Moussa {\it et al.},
                   Phys. Rev. B {\bf 54}, 15\ 149 (1996).

\bibitem{Nag97} S. Ishihara, M. Yamanaka, and N. Nagaosa,
                   Phys. Rev. B {\bf 56}, 686 (1997).

\end{references}
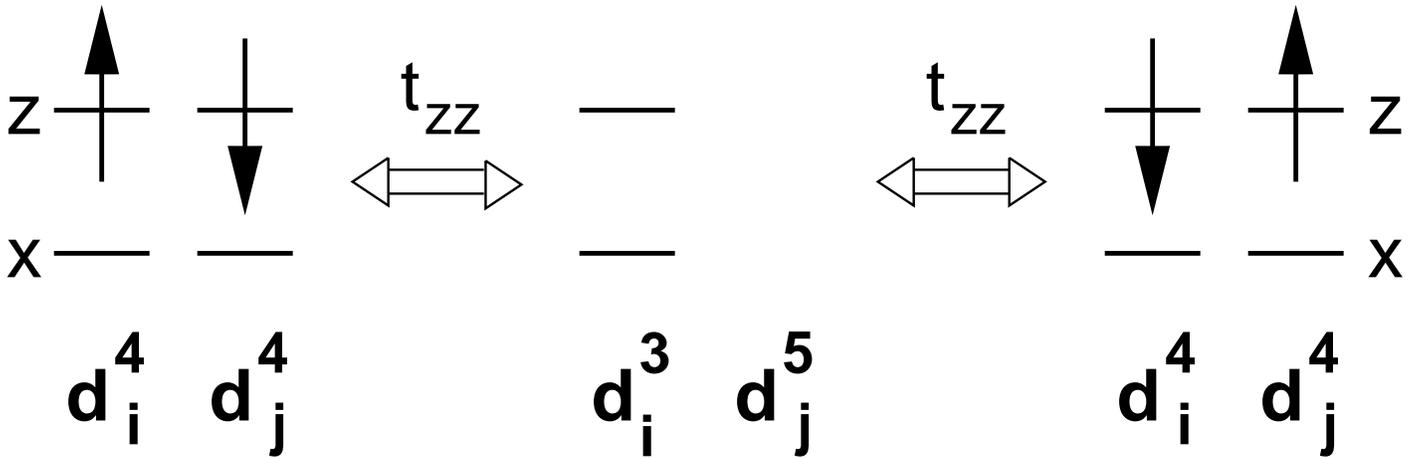
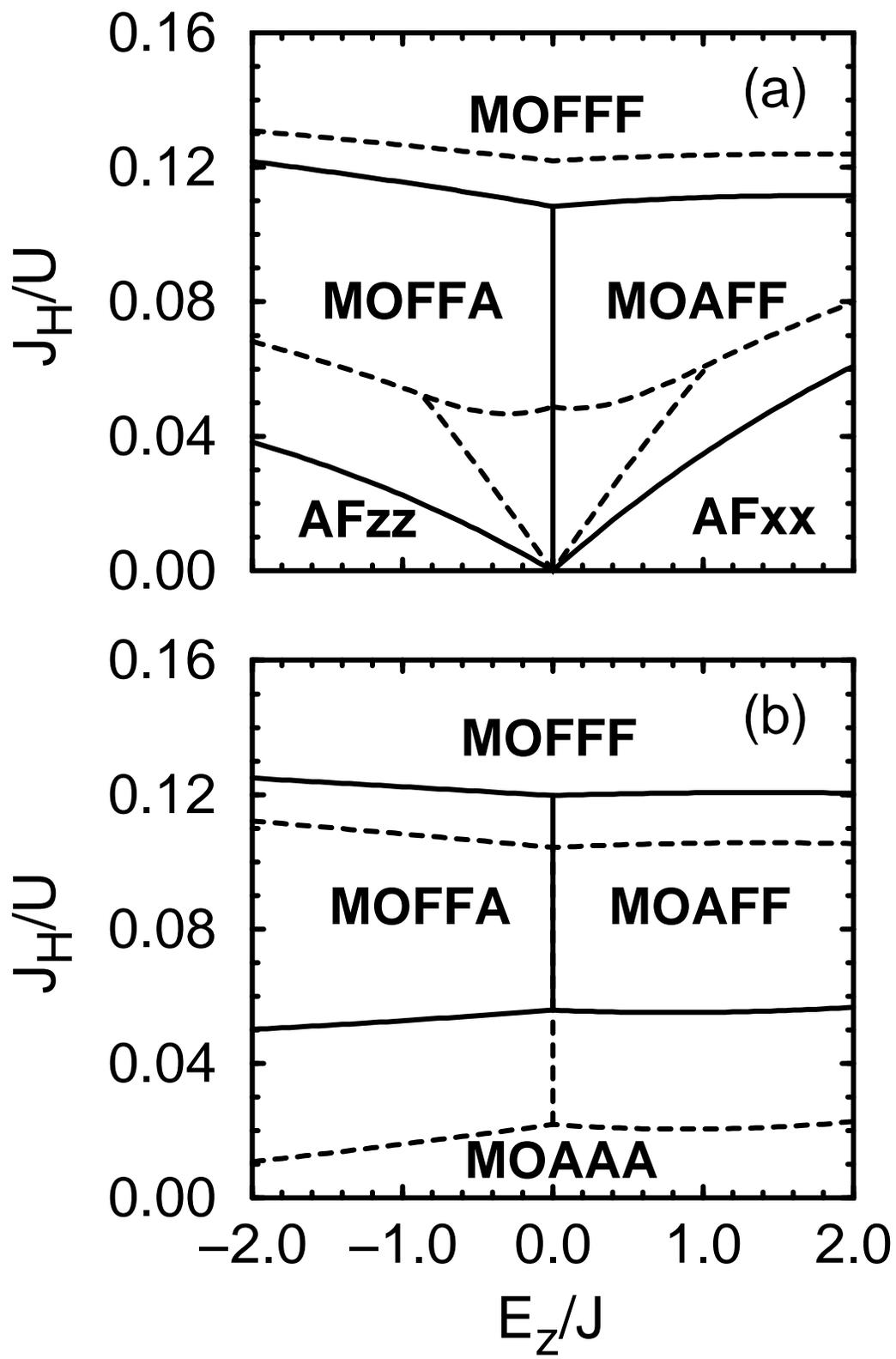
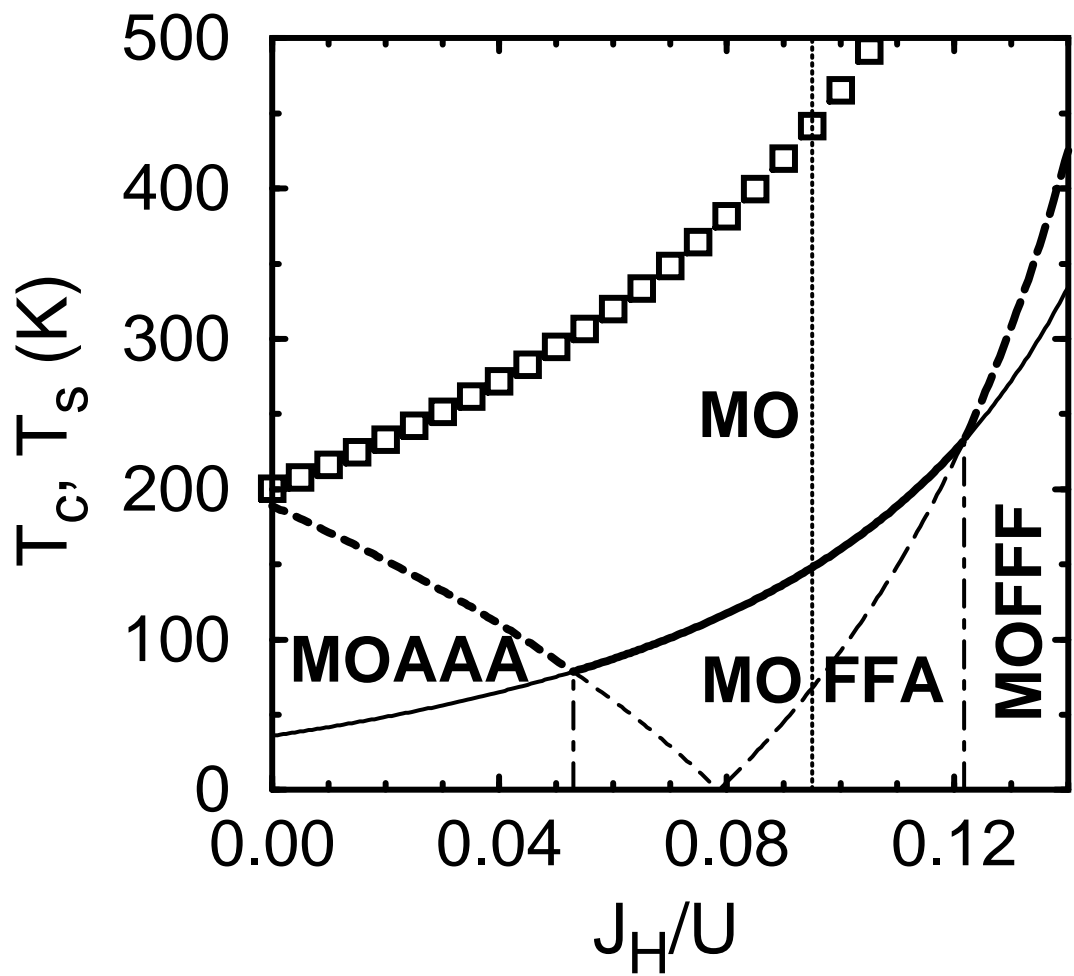
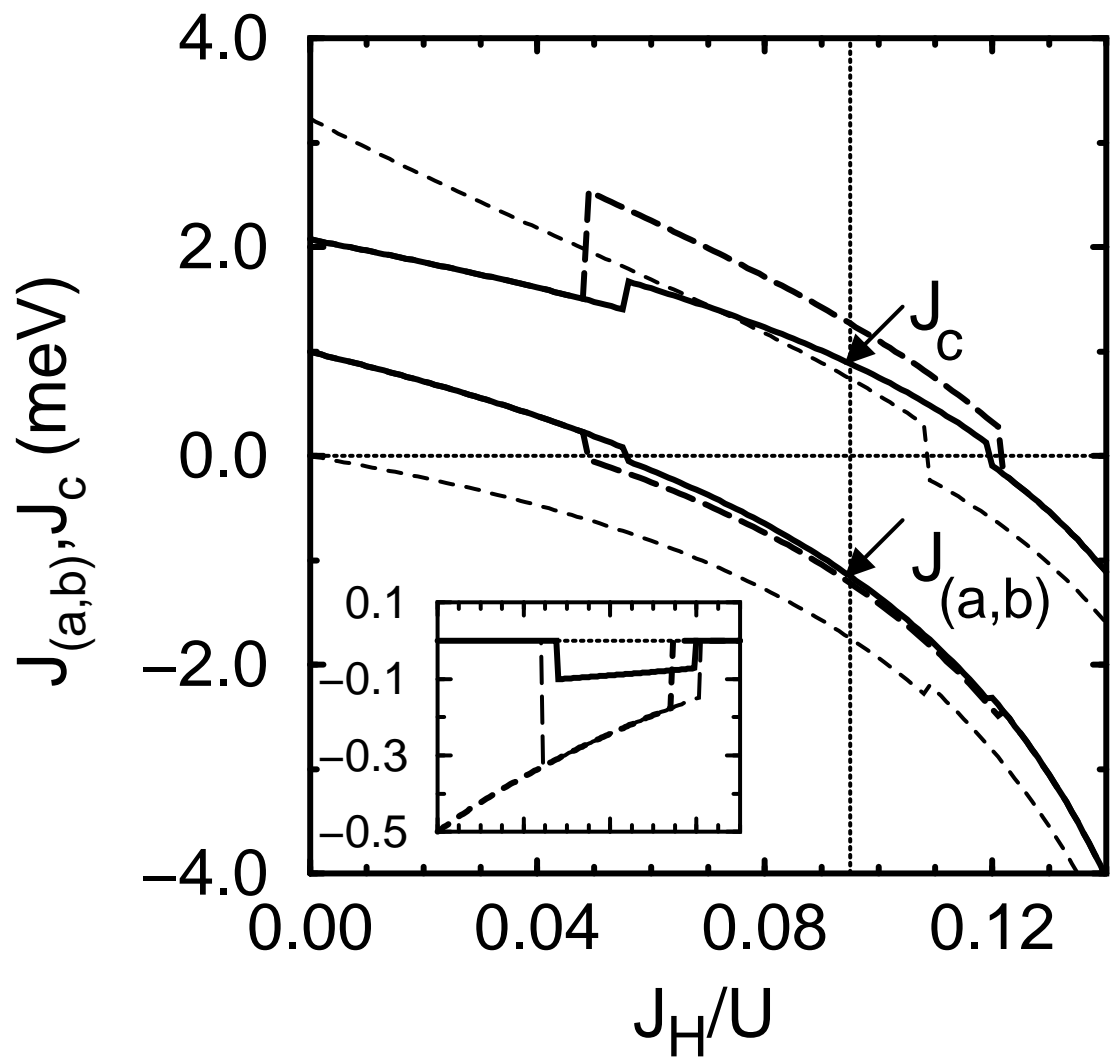
\end{document}